\documentclass[10pt,letterpaper,twocolumn]{article} 

\usepackage{ol2}
\usepackage[draft]{hyperref}
\usepackage{amsmath}

\begin{document}

\twocolumn[ 

\title{Phase locking of coupled lasers with many longitudinal modes}


\author{Moti Fridman, Micha Nixon, Eitan Ronen, Asher A. Friesem and Nir Davidson}

\address{Dept. of Physics of Complex Systems, Weizmann Institute of Science, Rehovot 76100, Israel}
\begin{abstract}

Detailed experimental and theoretical investigations on two
coupled fiber lasers, each with many longitudinal modes, reveal
that the behavior of the longitudinal modes depends on both the
coupling strength as well as the detuning between them. For low to
moderate coupling strength only longitudinal modes which are
common for both lasers phase-lock while those that are not common
gradually disappear. For larger coupling strengths, the
longitudinal modes that are not common reappear and phase-lock.
When the coupling strength approaches unity the coupled lasers
behave as a single long cavity with correspondingly denser
longitudinal modes. Finally, we show that the gradual increase in
phase-locking as a function of the coupling strength results from
competition between phase-locked and non phase-locked longitudinal
modes.
\end{abstract}

]

\noindent

 Phase locking of two coupled lasers operating with only
one longitudinal mode was investigated over the years
~\cite{RoySync, Glova, Fan, Fabiny}. It was shown theoretically
and experimentally that a simple relation exist between the
coupling strength that is needed for phase locking and the
frequency detuning between the lasers~\cite{Fabiny, FridmanOL,
VarditPRL}. While a sharp transition from no phase locking to full
phase locking when the coupling strength exceeds a critical value
is predicted, the experimental results revealed a gradual
transition, which could be explained by introducing noise to each
laser~\cite{Fabiny, VarditPRL}. For lasers with many longitudinal
modes it was shown that for strong coupling strength only common
longitudinal modes survive, leading to full phase
locking~\cite{LMinthe70, Shirakwa, shirakawaNoMore, Rothenberg}.
Yet, the detailed behavior of phase locking and the spectrum of
longitudinal modes as a function of the coupling strength between
coupled lasers were so far not reported.

Here we present our investigations and results on two coupled
fiber lasers, each operating with up to 20,000 longitudinal modes.
Specifically, we show how the phase locking between the two lasers
and their longitudinal mode spectrum vary as a function of the
coupling strength which is continually and accurately controlled
with polarization elements. We find  a gradual increase in the
number of longitudinal modes which are phase locked as the
coupling strength increases, leading to a gradual transition from
no phase locking to full phase locking without the need to
introduce noise. We support the experimental results with
calculations in which a modified effective reflectivity model is
exploited.

\begin{figure}[htb]
\centerline{\includegraphics[width=8cm]{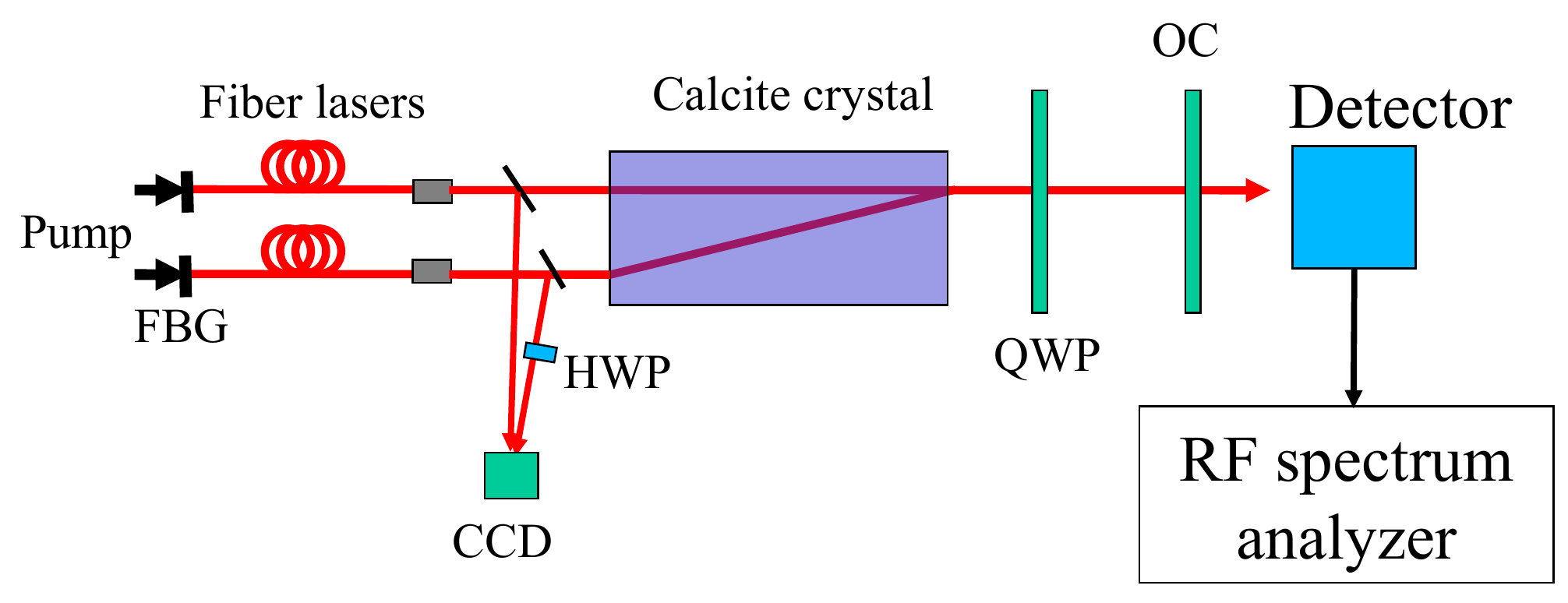}}
\caption{\label{configuration}Experimental configuration for
investigating the phase locking and the spectrum of longitudinal
modes of two coupled fiber lasers as a function of the coupling
strength. FBG - fiber Bragg grating. HWP - half wave plate. QWP -
quarter wave plate. OC - output coupler.}
\end{figure}

The experimental configuration for determining the phase locking
and the  spectrum of longitudinal modes for two coupled fiber
lasers as a function of the coupling strength between them is
presented in Fig.~\ref{configuration}. Each fiber laser was
comprised of a polarization maintaining Ytterbium doped fiber,
where one end was attached to a high reflection fiber Bragg
grating (FBG), with a central wavelength of $1064 nm$ and a
bandwidth of about $1nm$, that served as a back reflector mirror,
the other end attached to a collimating graded index (GRIN) lens
with anti-reflection coating to suppress any reflections back into
the fiber cores, and an output coupler (OC) with reflectivity of
$20\%$ common to both lasers. The lasers were pumped with $915nm$
diode lasers from the back end through the FBG. The two fiber
lasers were forced to operate in orthogonal polarizations by using
a calcite beam-displacer in front of a common output coupler and
the coupling strength $\kappa$ between the lasers was controlled
by an intra-cavity quarter wave plate (QWP).
$\kappa=\sin^2(2\theta)$, with $\theta$ the orientation of the QWP
with respect to the calcite main axes. The optical length of the
cavity of one fiber laser was $10m$ while the optical length of
the other was $11.5m$, so each fiber has $\sim 20,000$
longitudinal modes within the FBG bandwidth. The combined output
power was detected by fast photo detector which was connected to a
RF spectrum analyzer, to measure the beating frequencies and
determine the longitudinal mode spectrum at the
output~\cite{FridmanOL}. We also measured the phase locking
between the two fiber lasers by detecting the interference of
small part of the light from each laser with a CCD camera, and
determining the fringe visibility~\cite{VarditPRL}. The
longitudinal mode spectrum was measured at first when $\theta=0$
$(\kappa=0)$ and then sequentially repeated such measurement, each
after rotating the QWP by $1^\circ$ until we reached $45^\circ$
$(\kappa=1)$.

We developed a  model for calculating the distribution of
longitudinal modes and phase locking for the two coupled lasers.
For each laser, the effective reflectivity ~\cite{SigmanEffRef,
VarditEffRef} of its own reflection and the light coupled into it
from the other laser was calculated self consistently. The
longitudinal mode spectrum was then derived from the total
effective reflectivity of the two lasers. The effective
reflectivity resulting from the coupling to the other laser for
each laser can be shown to be,
\begin{equation}
R^{eff}_{1,2}=\left(1-r (1-\sqrt{\kappa})-\frac{r^2 \kappa
e^{\imath l_{2,1} k}}{1-r(1-\sqrt{\kappa})e^{\imath l_{2,1}
k}}\right)^{-1},
\end{equation}
where k denotes the propagation vector of the light, $\kappa$ the
coupling strength between the two lasers, $l_{2,1}$ the length of
each laser and $r$ the reflectivity of the output coupler. To
account for gain competition between the longitudinal modes we
used $r=0.55$ as a fitting parameter, rather than our experimental
value of $r=0.2$~\cite{VarditEffRef}. We then sum over the round
trip propagations to obtain the self consistent field for each
laser as,
\begin{equation}
\begin{array}{cr}
R^{eff}_{j}e^{\imath k l_j}+\left(R^{eff}_{j}e^{2\imath k
l_{j}}\right)^2
 +\ldots  = \left(1-R^{eff}_{j}e^{\imath k l_{j}}\right)^{-1},
\end{array}\label{eqR2}
\end{equation}
where $j=1,2$. Finally, the output laser field of the two coupled
lasers $R_{out}$, namely the amplitude of the longitudinal modes,
is obtained as a sum of the two self consistent fields, as
\begin{equation}
R_{out}=\left(1-R^{eff}_{1} e^{\imath k l_{1}
}\right)^{-1}+\left(1-R^{eff}_{2} e^{\imath k l_{2}}\right)^{-1}.
\end{equation}

\begin{figure}[htb]
\centerline{\includegraphics[width=8.3cm]{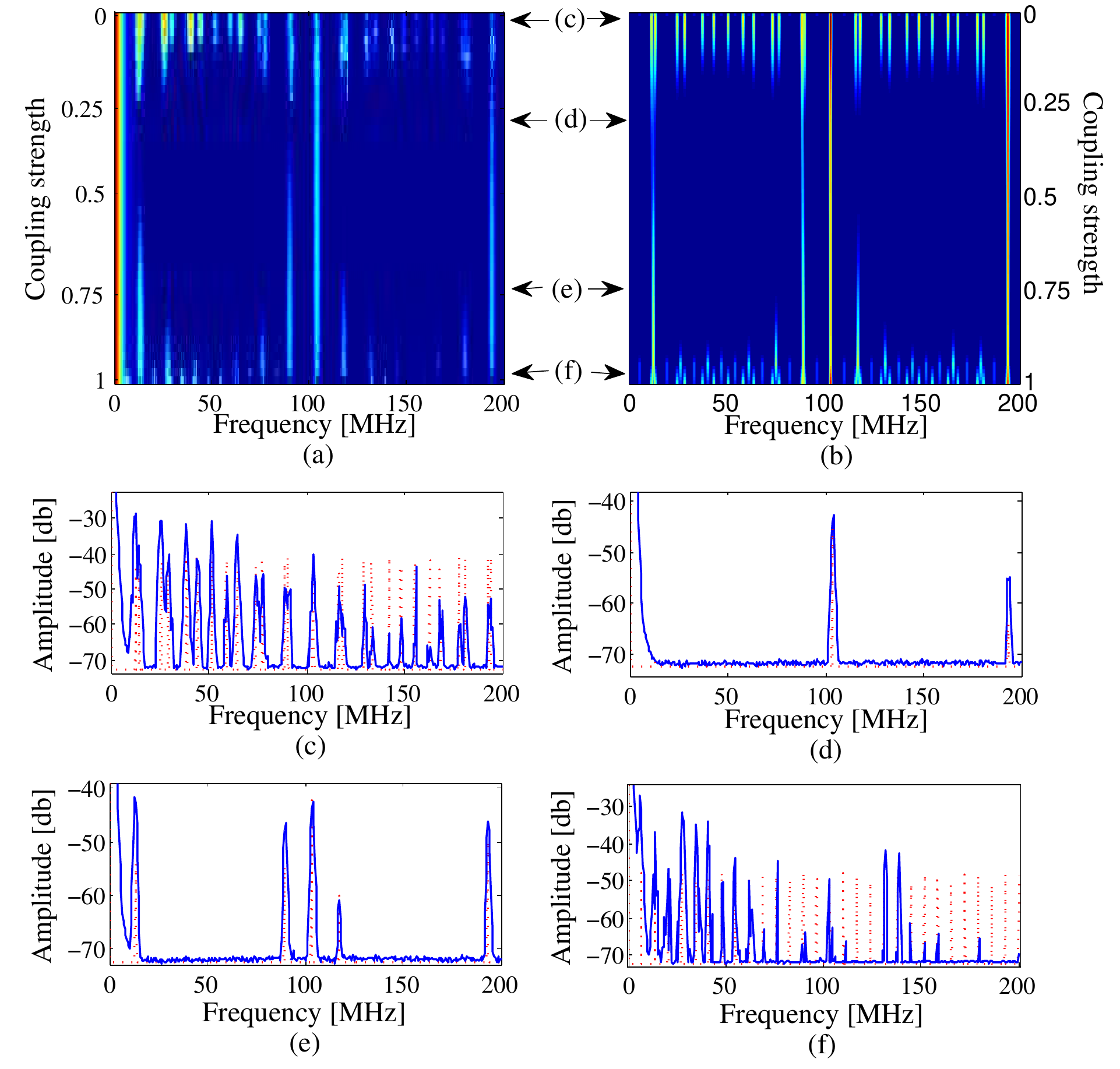}}
\caption{\label{Results}Experimental and calculated distributions
of longitudinal modes for two coupled lasers as a function of the
coupling strength $\kappa$. (a) Experimental results; (b)
calculated results; (c) $\kappa=0$; (d) $\kappa=0.28$; (e)
$\kappa=0.7$; (f) $\kappa=1$. Solid (blue) curves denote
experimental results and dotted (red) curves denote calculated
results. }
\end{figure}

Figure ~\ref{Results} shows the experimental and calculated
longitudinal mode spectrum as a function of coupling strength
$\kappa$ . Figure~\ref{Results}(a) shows the experimental results
of the longitudinal mode spectrum as a function of coupling
strength over $200MHz$ range, and Fig.~\ref{Results}(b) the
corresponding calculated results. Note that within our $1nm$ laser
bandwidths, the $200MHz$ range would be repeated many times. The
experimental and calculated results are also shown in greater
detail for four specific coupling strengths ($\kappa
=0,0.28,0.7,and \ 1$) in Figs.~\ref{Results}(c)-(f), respectively.
Without coupling (i.e. $\kappa=0$) two independent sets of
frequency combs exist simultaneously, one corresponds to the $10m$
long fiber laser ($15MHz$ separation between adjacent longitudinal
mode) while the other corresponds to the $11.5m$ long fiber laser
($13MHz$ separation), as also seen in Fig.~\ref{Results}(c). Each
7th longitudinal mode of the $10m$ long laser is very close to the
8th mode of the other, so they are essentially common longitudinal
modes. When $\kappa$ is increased from 0 to 0.3 the longitudinal
modes that are not common gradually disappear according to their
detuning while transferring their energy to the remaining ones via
the homogenous broadening of the gain. The longitudinal modes with
the larger detuning disappear first while the ones with smaller
detuning disappear for larger values of $\kappa$ and only the
common longitudinal mode remains, as also seen in
Fig.~\ref{Results}(d), indicating that at this coupling strength
there is full phase locking. As the coupling strength increases
above $0.3$ the longitudinal modes gradually reappeared, as also
seen in Fig.~\ref{Results}(e). The longitudinal modes with the
smaller detuning reappear first and the ones with larger detuning
reappear for larger values of $\kappa$. Finally, when $\kappa$
approaches unity, whereby all the light from one laser is
transferred to the other, new longitudinal modes appear in between
adjacent longitudinal modes,as also seen in Fig.~\ref{Results}(f),
corresponding to a single combined laser cavity whose length is
the sum of the two lasers.

Figure ~\ref{Results} reveals a good quantitative agreement
between the experimental and calculated results. In particular,
the observed gradual disappearance of non-common longitudinal
modes as the coupling is increased, their gradual reappearance
when the coupling is further increased and finally the doubling of
the frequency comb at near unity coupling strength are all
accurately reconstructed by our model.

\begin{figure}[htb]
\centerline{\includegraphics[width=8cm]{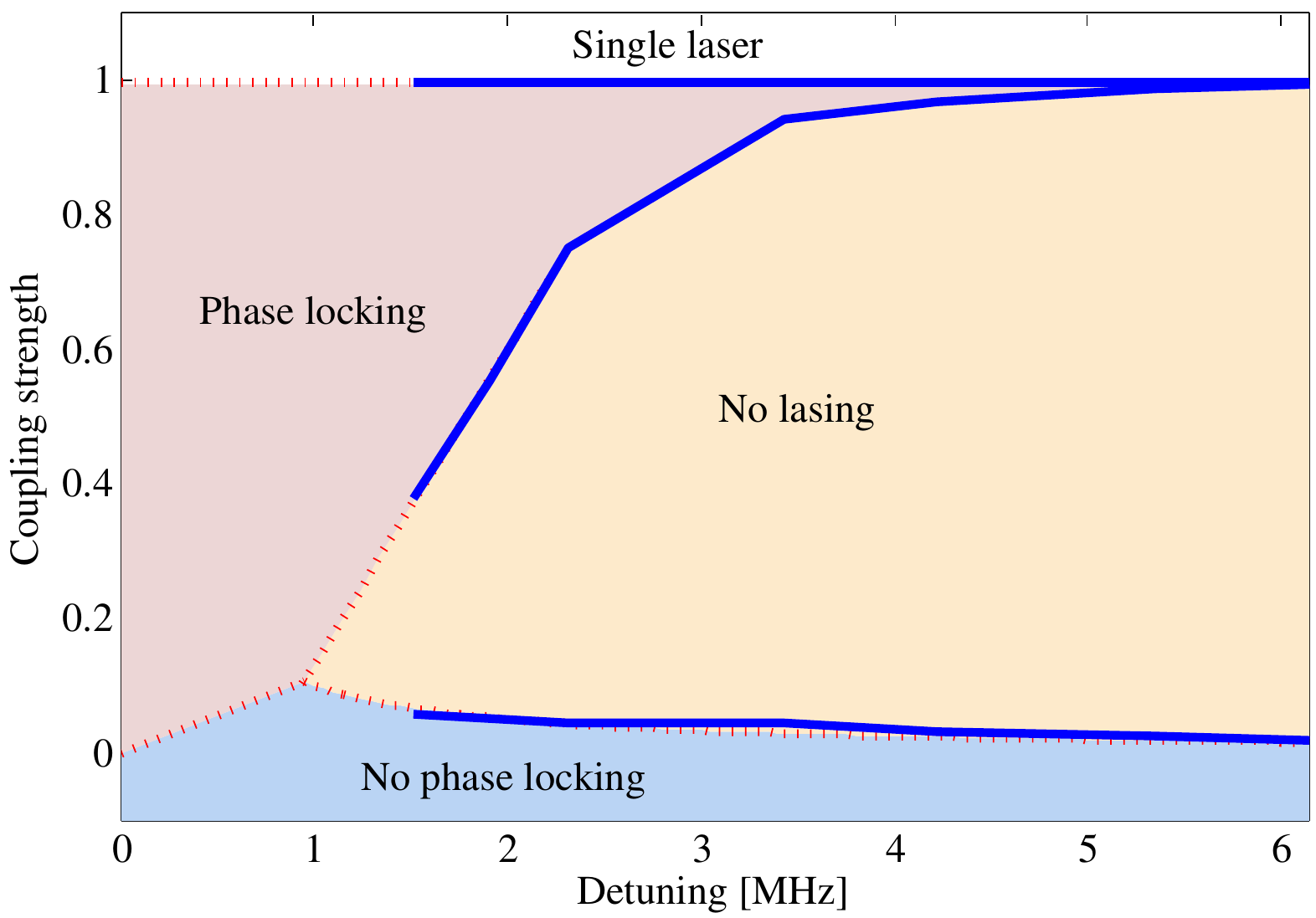}}
\caption{Experimental and calculated phase diagram of the
longitudinal modes behavior as a function of the coupling strength
between two coupled lasers and detuning between adjacent
longitudinal modes. Solid curves denote experimental results.
Dotted curves denote calculated results. \label{phasediagram}}
\end{figure}

The results of Fig.~\ref{Results} can be exploited to produce the
full phase diagram of the longitudinal modes behavior, as shown in
Fig.~\ref{phasediagram}. Here the behavior is presented as a
function of the coupling strength between the lasers and the
detuning between adjacent longitudinal modes. The phase diagram
includes four regions. In the first region of weak coupling the
longitudinal modes are not phase locked for any finite detuning
between them. In the second region of moderate to strong coupling
and small detuning, the longitudinal modes are phase locked,
indicating that the coupling is strong enough to overcome the
detuning. In the third region where the coupling strengths is not
sufficient to overcome the larger detuning between adjacent
longitudinal modes, these modes cannot lase. Finally, in the
fourth region, where the coupling strength approaches unity, the
two coupled lasers behave as a single and longer laser with denser
longitudinal modes. As seen, for detuning smaller than about
$1MHz$ there is a direct transition from no phase locking  to
phase locking as the coupling strength is increased. For larger
detunings the transition from  no phase locking to  phase locking
is interrupted by the no lasing region.

\begin{figure}[ht]
\centerline{\includegraphics[width=8cm]{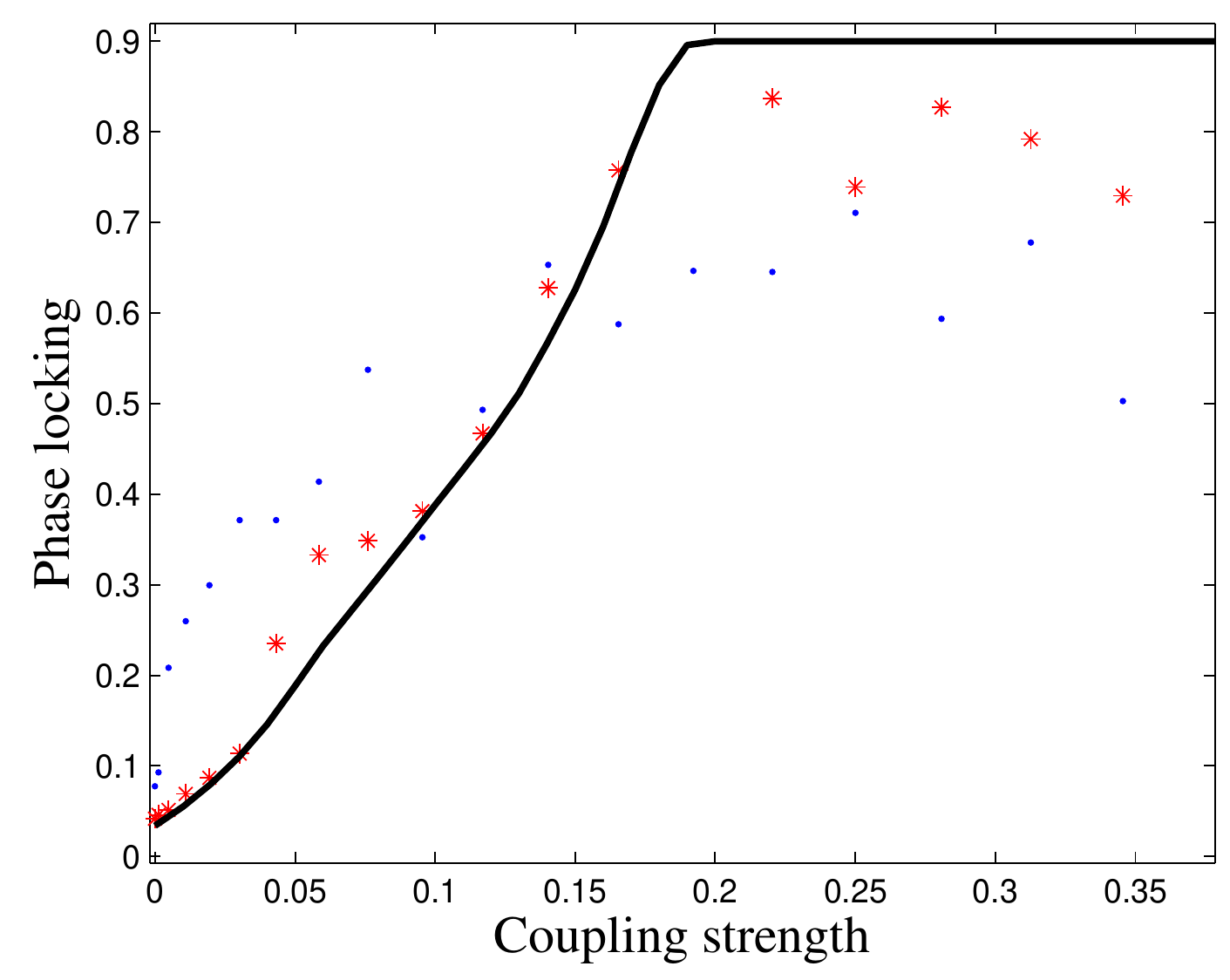}}
\caption{\label{phaselocking}Experimental and calculated phase
locking as a function of the coupling strength between two coupled
fiber lasers. Dots denote the directly measured phase locking;
stars denote the phase locking calculated from experimental
results in Fig~\ref{Results}(a); solid curve denotes the phase
locking calculated from the results in Fig.~\ref{Results}(b).}
\end{figure}

We also measured directly the phase locking (i.e. fringe
visibility) between the two coupled lasers as a function of the
coupling strength. The results, presented in
Fig.~\ref{phaselocking}, reveal a gradual increase in phase
locking as the coupling strength become stronger.
Fig.~\ref{phaselocking} also shows the ratio of the power in the
common longitudinal mode over that of all longitudinal modes for
each coupling strength, measured from the results of
Fig.~\ref{Results}(a). The good agreement between this ratio and
the direct measure of phase locking verifies that the gradual
increase in phase locking corresponds to the gradual disappearance
of the non-common longitudinal modes which are not phase-locked.
Finally, Fig.~\ref{phaselocking} also shows the ratio of the power
in the common longitudinal mode over that of all longitudinal
modes for each coupling strength, calculated from the results of
Fig.~\ref{Results}(b) which are also in good agreement with both
measurements.

To conclude, we presented how phase locking between two coupled
lasers which operate with many longitudinal modes depends on the
coupling strength and the detuning between the modes. We found
that there is a gradual transition from no phase locking to full
phase locking with increasing coupling strength due to the gradual
disappearance of longitudinal modes that are not phase locked. The
experimental results were confirmed with calculations using a
modified effective reflectivity model for coupled lasers that we
developed. Our results provide a fairly complete picture for
elucidating the behavior of the spectrum of the longitudinal modes
in coupled lasers and its effect on phase locking.

This research was supported in part by the USA-Israel Binational
Science Foundation.


\end{document}